\documentclass{JHEP3}
\usepackage{amsmath,amssymb,graphicx,subfigure,bm}
\usepackage{slashbox}
\newcommand\be{\begin{equation}}
\newcommand\ee{\end{equation}}
\newcommand\bea{\begin{eqnarray}}
\newcommand\eea{\end{eqnarray}}

\renewcommand\ap{\alpha^\prime}
\def\subsubsec#1{\bigskip\noindent{\begin{center}\textit{#1}\end{center}} \bigskip}

\def\d{{\partial}}
\def\db{{\bar{\partial}}}

\def\a{{\alpha}}

\def\d{{\partial}}
\def\db{{\bar\partial}}
\def\zb{{\bar z}}

\def\rhot{{\tilde{\rho}}}

\newcommand{\ket}[1]{| #1 \rangle}

\def\ie{{\it i.e.}}
\def\t#1{{\tilde{#1}}}

\newcommand\ymom{{{n_Y}}}
\newcommand\ymon{{{n_Y}}}
\newcommand\ywind{{{w_Y}}}
\newcommand\ny{{N_Y}}
\newcommand\tny{{\widetilde{N}_Y}}

\newcommand\tn{{\widetilde{N}}}

\def\cirpart#1{{\mathcal{Z}_{#1}(\tau)}}
\def\cirpartB#1{{\mathcal{Z}_{#1}}}
\def\cirab#1#2{{\mathcal{Z}_R}^{#1}_{\ \ #2}}
\def\cirNnw#1#2{{\mathcal{Z}_{NR}}[#1|#2]}
\def\cir4nw#1#2{{\mathcal{Z}_{4R}}[#1|#2]}

\def\zta#1#2{{\zeta_{#1#2}}}
\def\vt#1#2{{\vartheta_{#1#2}}}
\def\eqn#1{(\ref{#1})}
\def\f#1#2{\frac{#1}{#2}}

\newcommand\q{{\mathcal{Q}}}

\newcommand\qq{{
Q\biggl[\
\begin{matrix}
a \\ b
\end{matrix}
\ \bigg|\
\begin{matrix}
\alpha \\ \beta
\end{matrix}
\ \biggr] }}

\newcommand\qqt{{
Q(\tau)\biggl[\
\begin{matrix}
a \\ b
\end{matrix}
\ \bigg|\
\begin{matrix}
\alpha \\ \beta
\end{matrix}
\ \biggr] }}

\def\Q#1#2#3#4{
Q\biggl[\
\begin{matrix}
#1 \\ #3
\end{matrix}
\ \bigg|\
\begin{matrix}
#2 \\ #4
\end{matrix}
\ \biggr] }

\newcommand\mns{{\text{NS}}}

\newcommand\bS{{\begin{sideways}}}
\newcommand\eS{{\end{sideways}}}

\newcommand\rtau{{\tau_1}}
\newcommand\itau{{\tau_2}}

\def\va#1#2{$\begin{array}{c} #1 \\ #2 \end{array}$}
\def\vb#1#2{$\begin{array}{c} #1 \\ #2 \\ \ \end{array}$}

\title{Nongeometry, Duality Twists, and the Worldsheet}

\author{Alex Flournoy \\
  Department of Theoretical Physics\\
  Research School of Physical Sciences and Engineering \\
  The Australian National University, Canberra ACT 0200, Australia \\
  Email: \email{Alex.Flournoy@anu.edu.au}}

\author{Brook Williams\\
Institute for Theoretical Physics \\
University of Amsterdam 1018 XE Amsterdam, the Netherlands \\
Email: \email{brook@science.uva.nl}}

\abstract{In this paper, we use orbifold methods to 
construct nongeometric backgrounds, and argue that they
correspond to the spacetimes discussed in \cite{dh,wwf}.
More precisely, we make explicit through several examples
the connection between interpolating orbifolds
and spacetime duality twists.
We argue that generic nongeometric backgrounds arising from duality twists
will not have simple orbifold constructions and
then proceed to construct several examples which do have consistent worldsheet descriptions.}

\preprint{hep-th/0511126 \\
ITFA-2005-47}

\begin{document}

\section{Introduction}

Recently there has been a lot of work in which symmetries and dualities
have been exploited in order to construct string theory backgrounds
\cite{dh,wwf,hmw,hulltwist1,italy,hulltwist2}.  In particular, these
backgrounds may be described as fibrations in which the fibers
undergo non-trivial monodromies around various points or cycles in
the base. It has been argued that such a description is quite
generic \cite{syz}. The basic idea dates back to the work of
Scherk-Schwarz \cite{schkschz} who, in the context of classical
supergravity, imposed nontrivial boundary conditions to give masses
to lower dimensional fields. Adopting these ideas to string theory
brought two changes.  First, the global symmetries of the low energy
effective action which allow one to impose nontrivial boundary
conditions are broken to discrete subgroups.  Secondly the (now
discrete) symmetry group may be enlarged to include the full stringy
duality group. In these cases the fibers are said to undergo
``duality twists''\cite{dh}. Related ideas have been around for some
time. In \cite{umanifolds} for example, backgrounds are constructed
by allowing for monodromies coming in the U-duality group. There are
many reasons to be interested in these \lq\lq twisted''
compactifications. For starters, it is often the case that moduli
are projected out by the boundary conditions. Moreover, they are
closely related to compactifications with H-flux,  supersymmetric
non-K\"ahler vacua, and nongeometric
backgrounds\cite{italy,kstt,beckerdas,stw,su31,su32,su33,ck,gh-j}.

Nongeometric backgrounds are of particular interest since relatively
little is known about them. Moreover, this lack of knowledge touches
on some of the key questions facing any theory of quantum gravity.
What new structures will exist that were not present in the
classical point particle theory? What sort of framework will be
needed in order to naturally accommodate such structures? There has
been progress recently addressing the larger question of how to
naturally accommodate some of these structures: the development of
\lq\lq Generalized Complex Geometry''\cite{gcg,marco} naturally
incorporates the B-field, and Hull's \lq\lq doubled
formalism'' \cite{hull}, which is closely related to the \lq\lq
correspondence spaces" of Bouwknegt et al \cite{bem1,bem2}, gives
nongeometric spaces a (higher dimensional) geometric description.

This paper will address the first question by focusing on the worldsheet 
construction of a particular class of nongeometric
backgrounds. In particular we look at backgrounds which come from
torus fibrations over a torus base where the torus fiber undergoes a
monodromy lying in the perturbative duality group
$O(d,d;\mathbb{Z})$.  Following \cite{wwf} we'll refer to these as
monodrofolds, although they have also been called T-folds in the
literature \cite{hull}\footnote{We take the term \lq\lq T-fold'' to
be specific to monodromies coming from the T-duality group, where
the monodromies in a \lq\lq monodrofold'' may be more general.  This
being said, throughout this paper the terms will be
interchangeable.}. In \cite{wwf} it was shown from the spacetime
perspective that the equations of motion forced many of the fields
to be fixed under the action of the imposed monodromy, projecting
out moduli which would arise in a traditional compactifications.
Similar conclusions were reached in \cite{dh}. Although finding the
spacetime description for these backgrounds is somewhat
straightforward, the worldsheet construction is often a difficult
task. In fact, due to their asymmetric nature, quantum consistency
is not guaranteed.

As suggested in \cite{dh,wwf}, duality twists correspond to
interpolating orbifolds. An interpolating orbifold is just an
orbifold where the group action integrates a translation along one
coordinate with nontrivial transformations of the remaining
coordinates or fermions. For example, one might consider orbifolding
a boson $X$ by a reflection while orbifolding a second boson $Y$ by
a $2\pi R$ shift, \be\label{intrexp} (X \rightarrow -X \ , \ \ Y
\rightarrow Y + 2\pi R ). \ee  This should be distinguished from the
product orbifold\footnote{The product orbifold corresponds to
separately compactifying Y and orbifolding by a reflection in X.}
which would contain 3 distinct actions, \bea (X\rightarrow -X , Y
\rightarrow Y) ; (X \rightarrow X , Y \rightarrow Y + 2\pi R) ;
(X\rightarrow -X , Y \rightarrow Y + 2\pi R) \ . \eea As one might
expect, orbifolding by \eqn{intrexp} corresponds in spacetime to a
circle fibration in which the fiber undergoes a reflection as it
encircles the base. The name \lq\lq interpolating orbifold'' comes
from the fact that as $R \rightarrow \infty$ the orbifold returns
to the unorbifolded theory, while for $R \to 0$ the theory is
indistinguishable from the product orbifold in the same limit. Such orbifolds can then
be used to find theories which, for example, interpolate between
backgrounds with different amounts of
supersymmetry\cite{rohm,it,gv}.

A shift orbifold alone is modular covariant\footnote{This is
shown in appendix \ref{ap:shift}.}. Namely, defining
${\mathcal{Z}_R}^a_{\ \ b}$ to be the contribution to the shift partition
function coming from $a$ twists and $b$ insertions,
\be
{\mathcal{Z}_R}(\tau + 1)^a_{\ \ b} = {\mathcal{Z}_R}(\tau)^a_{\ \ b-a} \ \
\mbox{and} \ \ 
{\mathcal{Z}_R}(-1/\tau)^a_{\ \ b} = {\mathcal{Z}_R}(\tau)^b_{\ \ -a} \ .
\ee 
Modular covariance is a stronger condition than modular invariance,
but is often much easier to check.
It is easy to convince oneself that combining two sets of
sectors which are individually modular covariant will always lead to
a modular covariant, and in turn a modular invariant, theory. 
This leads to a great simplification in these constructions, since
the
modular invariance of an interpolating orbifold will be guaranteed 
if the pure orbifold, the orbifold without the coordinate
shift, is modular covariant.

In section \ref{sec:circ} we will discuss the simplest example of a
nongeometric monodromy: a circle fibered over another circle in
which the fiber undergoes T-duality. This example is not modular
invariant, but it serves as a useful illustration of where the
consistency can fail. In section \ref{sec:wwf} we move on to the
case of a 2-torus fiber, making contact with the work in \cite{wwf}.
We will check that the orbifold construction does indeed correspond
to the spacetime constructions discussed in \cite{wwf}, and then, in
order to check quantum consistency, we will explicitly construct the
partition function and check modular invariance. As will be shown,
these particular asymmetric orbifolds are dual to symmetric
orbifolds, connecting these seemingly exotic backgrounds to
more traditional compactifications. In section \ref{sec:het}, we'll
consider a more general torus fibration in the context of the
heterotic string.  It will be shown that, satisfying certain
constraints on dimensionality, one can construct modular invariant
interpolating asymmetric orbifolds which are truly, in the sense
that they do not have a geometric dual, nongeometric.  As a final example,
in section \ref{sec:shift} we construct
interpolating orbifold realizations of
monodrofolds involving modular shifts.  In doing so we give a simple
example of how to construct interpolating orbifolds when the corresponding
pure orbifold is itself a product orbifold, and discuss the subtleties which 
subsequently arise.

\section{The circle twist and its discontents}\label{sec:circ}

Let's consider the most trivial nongeometric twist. Namely, a circle
(at the $SU(2)$ radius) fibered over another circle, where the fiber
circle undergoes a T-duality transformation.  From the worldsheet
perspective this can be understood as an orbifolding by the action
\bea
X_R &\rightarrow& -X_R \nonumber\\
X_L &\rightarrow& \ ~X_L \nonumber \\
Y~ &\rightarrow& Y + 2\pi R \ . \label{circorb} \eea Here $X/Y$ is
the embedding coordinate corresponding to the fiber/base circle.

As discussed in the introduction, the quantum consistency of
\eqn{circorb} is dependent upon the modular covariance of the
orbifold without the shift, \bea\label{chiralreflect} X_R
\rightarrow -X_R \ , \ \ X_L \rightarrow X_L \ . \eea Its not
hard to see that the theory corresponding to \eqn{chiralreflect} is not modular invariant, much
less modular covariant. Consider the partition function for a
symmetric reflection orbifold, \be\label{symmrefl} Z_{S^1/R} =
{\f 1 2}Z_{S^1} +|\f {\eta(\tau)} {\vt 10 (\tau)}|
+| \f {\eta(\tau)} {\vt 01 (\tau)}| +|\f {\eta(\tau)} {\vt 00 (\tau)}|.
 \ee
Here $\eta(\tau)$ are the Dedekind eta functions and 
$\vt \alpha \beta (\tau) \equiv \theta_{\alpha \beta}(0,\tau)$, where 
$\theta_{\alpha \beta}(\nu,\tau)$ are the 
theta functions with characteristics.
It is straightforward to show that, up to phases, the $\eta/\vt
\alpha \beta$ transform into one another under modular
transformations.  Since \eqn{symmrefl} contains the absolute value
of these terms the phases cancel and the entire expression is
modular invariant. Now consider the asymmetric reflection
\eqn{chiralreflect}, which we will also refer to as a chiral
reflection. From what we've learned in the symmetric case one might
naively write down the expression, 
\be\label{asymmrefl}
Z_{S^1/T}^{\rm NAIVE} = {\f 1 2}Z_{S^1} +
 (\f {\eta(\tau)}{\vt 10 (\tau)})^{\f 1 2}Z^r_{S^1} 
+(\f {\eta(\tau)}{\vt 01 (\tau)})^{\f 1 2}Z^r_{S^1} 
+(\f {\eta(\tau)}{\vt 00(\tau)})^{\f 1 2}Z^r_{S^1}
 \ee
where $Z^r_{S^1}$ is the contribution to the $S^1$ partition
function from the right movers. Though we have not explicitly shown
what $Z^r_{S^1}$ is, it should be clear that the trivial cancellation
of phases which took place in \eqn{symmrefl} will not occur in
\eqn{asymmrefl}. With a little work one can show that
\eqn{asymmrefl} is not modular invariant.

With the simplest possible asymmetric orbifold failing one may begin
to worry.  However, there are many examples of modular invariant
asymmetric orbifolds: Orbifolds involving chiral reflections (and
chiral shifts) are precisely the subject of \cite{adp}. There it was
shown that not only was the above expression not modular invariant,
it was the wrong expression altogether.  As discussed in
\cite{adp}, our intuition on how to construct this orbifold breaks
down due to a phase ambiguity that arises when splitting up the
rotation $R(\theta)$ into a left/right-handed piece; $R(\theta)
=R_L(\theta)R_R(\theta)$.  In particular they show that the chiral
reflection is 4th (rather than 2nd) order, and that one must be in
$4m\in 4\mathbb{Z}$ dimensions if one hopes to have a modular
invariant theory. There are also more elaborate examples.  For
example in \cite{Erler,KS,Blum} it was shown that starting with a
$T^4$ at the $SO(8)$ point and combining chiral reflections of four
coordinates with half-shifts along each $T^4$ cycle results in a
modular invariant theory.

\subsubsec{Interpolating Orbifolds, Wilson lines and Quantum
Consistency}

Before moving on to modular invariant orbifolds we would like to
address one final issue concerning \eqn{circorb}.  It might seem
surprising that \eqn{circorb} is not consistent, since it looks very
much like turning on a Wilson line. To review:  At the self-dual
radius our symmetry group is enhanced from $U(1)_L \times U(1)_R$
to $SU(2)_L \times SU(2)_R$. The generators for the $SU(2)_L$ are,
\bea
j^+_X &=& \exp \bigl(+i 2 X_L /\sqrt{\ap} \bigr) \\
j^-_X &=& \exp \bigl(-i 2 X_L /\sqrt{\ap}\bigr) \\
j^3_X &=& \d X \ , \eea and similarly for $SU(2)_R$.  One is then
able to form vertex operators which lead to marginal deformations of
our worldsheet theory.  One may think of these deformations in terms
of Wilson lines since, \bea \int e^S &\rightarrow& \int
e^S\exp\bigl( \int d^2\sigma (\d Y \alpha \cdot \bar{j}_X
                                               +   \bar{\d} Y \beta \cdot j_X)\bigr) \\
         &=&  \int e^S \exp\bigl( \int d^{10}X^\mu (\bar{A}_\mu + A_\mu) \bigr)  \ .
\eea Where $\alpha^a,\beta^a$ are constant vectors whose indices run
over $a = +,-,3$, 
\be A_Y \equiv \int dz ~\beta \cdot j_X \ \
{\rm and} \ \ \bar{A}_Y \equiv \int d\bar{z} ~\alpha \cdot
\bar{j}_X \ , 
\ee 
and all other components of $A_\mu$ vanish.
To see how these Wilson lines may affect boundary
conditions consider deforming the theory by the operator,
\be\label{tohdef} \f 1 4 \d Y\bar{\d}Y + \f 1 2 (\d Y\bar{\d}X + \d
X\bar{\d}Y) \ . \ee This is of course nothing more than a metric
deformation.  To be precise, if we started with $X,Y$ being the
cycles of a $T^2$ with $\tau = i$, the deformed theory would have
$\tau = 1/2 + i$.  This skewed torus, is then just a $S^1$ fibered
over a $S^1$ where the fiber shifts half way around as it circles
the base.

Let's now return to our original question: is there a Wilson line
which we can turn on which corresponds to imposing a chiral rotation
as a boundary condition? Using our $SU(2)$ symmetry we can ask an
equivalent, but more transparent question.  Is it possible to turn
on an operator which corresponds to imposing a chiral shift boundary
condition? Using what we've learned from \eqn{tohdef} we can see
that such an operation should involve $\bar{j^3}$.  For instance one
might suggest $\d Y\bar{\d}X + \bar{\d}Y\bar{\d} X$, but one
must not forget that we are looking for marginal deformations, i.e.
weight $(1,1)$.  The $\bar{\d}Y\bar{\d} X$ piece breaks conformal
invariance.  A more plausible suggestion would be $(1/4) \d
Y\bar{\d}Y + \d Y\bar{\d}X$.  However, it is not difficult to show
that this simply corresponds to a symmetric background with $\tau =
\rho = 1/2 + i$.  In fact any linear combination of marginal
operators involving $\d Y, \bar{\d} Y, j^3 , \bar{j}^3$ will amount
to nothing more than skewing the torus with a constant B-field.
Moreover, since we are only interested in shifts, the operators
$j^\pm,\bar{j}^\pm$ do not play a role. Our claim is then, although
there are elements of $SU(2)_L\times SU(2)_R$ which correspond to
chiral shifts or rotations, there are no $(1,1)$ operators which
correspond to imposing these boundary conditions. Moreover, we
believe, though we have certainly have not proved it here, that all
marginal deformations of the world sheet theory are geometric, and
therefore there is no way to deform from any geometric theory into
any truly nongeometric theory, and vice versa.

\section{Symmetric and pseudo asymmetric orbifolds}\label{sec:wwf}

We would now like to turn our attention to backgrounds which are
modular invariant. There are two classes which we know will work:
backgrounds which are dual to geometric backgrounds, and backgrounds
which do not have a geometric dual but come from a consistent asymmetric
orbifold.  Distinguishing truly nongeometric compactifications from
those which are dual to geometric compactifications can sometimes be
tricky. However, the spacetime picture can help.

Consider a $T^2$,  with a complex structure $\tau$ and the kahler modulus
$\rho$, fibered over a
circle.  The symmetry group of the torus is $SL(2,\mathbb{Z})_\tau
\times SL(2,\mathbb{Z})_\rho$, where the  $SL(2,\mathbb{Z})_\tau$ is
the geometric symmetry group of the torus and
$SL(2,\mathbb{Z})_\rho$ comes from T-duality and B-field shifts. If
one looks at a monodromy $\rho \rightarrow -1/\rho$ the background
naively looks nongeometric. However, T-dualizing on one of the torus
cycles exchanges $\rho$ and $\tau$, and therefore, in its T-dual
picture, the background is geometric.  We will refer to such
backgrounds as being pseudo asymmetric (nongeometric). Now consider the
monodromy $\tau \rightarrow -1/\tau,\ \rho \rightarrow -1/\rho$.
Because of the symmetry of the $\rho$ and $\tau$ monodromies,
anything we do to make the nongeometric monodromy geometric will
also make the geometric monodromy nongeometric.  Such a background
does not have a dual geometric interpretation, and will be referred
to as truly asymmetric (nongeometric).  In the rest of this section 
we will discuss a symmetric orbifold and its pseudo asymmetric dual.
In the next section we'll return to the discussion of truly asymmetric 
orbifolds\footnote{Here we have argued that there is no dual coming from the 
perturbative string duality group, one should really consider the entire
U-duality group.  This being said we don't suspect that any geometric dual 
will exist.}.

The focus of \cite{wwf} is on the spacetime construction for $T^2$
fibrations over a $T^2$ base.  In particular it looks at backgrounds
where the $T^2$ fiber undergoes a nongeometric twist. The
supersymmetry transformations require \bea\label{sugra} \db \rho =
\db \tau = \d \db (\ln \rhot_2 - \ln \rho_2 - \ln \tau_2 ) = 0 \ .
\end{eqnarray}
Here $\tau~ (\tilde{\tau})$ and $\rho~ (\tilde{\rho})$ denote the
complex structure and K\"ahler moduli of the fiber (base),
respectively. Generically such solutions preserve (1,0)
supersymmetry in six dimensions.  For the type IIA theory, solutions
with constant $\rho$ preserve (1,1) supersymmetry and solutions with
constant $\tau$ preserve (2,0) supersymmetry. It follows from
(\ref{sugra}) that there is a nontrivial backreaction on the base
which forces the corresponding fiber moduli to be periodic.  If one
then imposes nontrivial (not periodic) boundary conditions the
moduli must take values which are fixed under the action of the
monodromy. For completeness, the details of this argument as
originally presented in \cite{wwf} are repeated in appendix
\ref{ap:constant}.

As an example consider type II A with the monodromy:
\begin{eqnarray}\label{wwfmonodromy}
\rho \rightarrow \rho \ ,  \ \tau \rightarrow -1/\tau \ \ \
(\tilde{\theta}^1-cycle)& \cr \rho \rightarrow -1/\rho \ , \  \tau
\rightarrow \tau \ \ \ (\tilde{\theta}^2-cycle)& \ .
\end{eqnarray}
It follows that $\tau$ and $\rho$ are fixed to the values $\rho =
\tau = i$. The fiber moduli coming from the NSNS sector are
projected out by the monodromy
(\ref{wwfmonodromy}).\footnote{Additionally, half of the RR moduli
are lifted \cite{wwf}.}

The lifting of moduli with boundary conditions is the central point
of these constructions. If one were only to consider the values of
the 10-dimensional metric and B-field they would see nothing more
than a flat metric with vanishing B-field.  However,
\eqn{wwfmonodromy} contains additional information, namely, the
boundary conditions of the fluctuations about this background. These
boundary conditions manifest themselves in the field content of the
dimensionally reduced theory. In particular, for
(\ref{wwfmonodromy}) there are no massless six-dimensional fields
coming from NSNS moduli of the $T^2$ fiber. It is worth noting that
even in traditional compactifications one is specifying (periodic)
boundary conditions. However there is no reason from a string's perspective
to place preference on periodic boundary conditions over other ones.
It is certainly an interesting question to ask if there is some
dynamical, thermodynamic or other way in which a string could
``choose'' one boundary condition over another.

The implications of nontrivial boundary conditions
on the low energy spacetime physics was discussed in detail in \cite{wwf}.
The rest of this paper discusses these same
backgrounds from a worldsheet perspective. In constructing the
worldsheet theory we will be able to study the full spectrum, both
massless and massive, and check quantum consistency.

As discussed above, we will construct these backgrounds in terms of
interpolating orbifolds. The nongeometric nature of these
compactifications will be reflected in the fact that the orbifolds
themselves are asymmetric. A great deal of work has gone into the
construction of consistent asymmetric
orbifolds\cite{nsvi,nsvii,adp}.  It is a very subtle procedure which
one would like to avoid if at all possible. Below we will discuss a
pseudo asymmetric orbifold. Since modular invariance is guaranteed
from the dual symmetric orbifold, the discussion is greatly
simplified. In the next section we will move on to a truly
asymmetric example.

Let us first consider a spacetime in which the $T^2$ fiber undergoes 
a monodromy $\tau \rightarrow -1/\tau$ as it traverses a circle
of radius $4R$, $S^1_{4R}$. From the
worldsheet perspective this corresponds to orbifolding the theory on a $T^2 \times
S^1_{4R}$ by the action, 
\bea
X^1 &\rightarrow& -X^2 \nonumber\\
X^2 &\rightarrow& X^1 \nonumber \\
Y~   &\rightarrow& Y + 2\pi R \ . \label{symorb} 
\eea 
Here $X^1$ and $X^2$ are the
periodic bosons of the $T^2$ and $Y$ is the boson coming from the
$S^1_{4R}$ (a circle of radius $4R$). The dual pseudo nongeometric
monodromy is $\rho \rightarrow -1/\rho$. This corresponds to two
T-dualities and a 90 degree rotation. The orbifold action is then
\bea
X^1_L \rightarrow -X^2_L \ &,&\ \ X^2_R \rightarrow -X^1_R \nonumber \\
X^2_L \rightarrow X^1_L \ &,&\ \ X^1_R \rightarrow X^2_R \nonumber\\
Y &\rightarrow& Y + 2\pi R \ . \label{pasymorb} 
\eea 
For the type II
string, modular invariance is guaranteed since
(\ref{symorb},\ref{pasymorb}) is a symmetric/pseudo-asymmetric
orbifold. For completeness we will write down the partition
functions below.  However, we should first check that orbifolding by
(\ref{symorb}) does indeed project out the same moduli as the
corresponding monodrofold.

\subsubsec{Moduli Fixing}

From the worldsheet perspective giving a mass to fluctuations in
the metric and/or B-field corresponds to projecting out particular
massless closed string states.  We'll discuss the fate of these
states below, but first we must build our Hilbert space. As 
usual, we quantize the system around a flat background (namely
$ds^2= \eta_{\mu \nu} dX^\mu dX^\nu$, $B = 0$) and then build closed
string states by acting repeatedly with the operators $\a^\mu_{-n}
\tilde{\a}^\nu_{-m}$\footnote{Here we are discussing the bosonic
string.  In the Type II string there are slight modifications to the
details of the argument, the bosonic oscillators are now world sheet
fermions $\psi^{NS}_{-n \ \mu}, \tilde{\psi}^{NS}_{-n \ \nu}$ for
instance, but the end result is the same.}. The states are labeled
by the occupation numbers $(N_i,\tn_i)$ and zero modes $(n_i,w_i)$
associated with the fields $X^i$ and $Y$, namely, \be \ket{
N_1,\tn_1,n_1,w_1;N_2,\tn_2,n_2,w_2;
      \ny,\tny,\ymom,\ywind ; \cdots } \ .
\label{st}
\ee
The $\cdots$ represents the eigenvalues coming from the other
fields in the theory; they will henceforth be left out.  Under
(\ref{symorb}) we find that (\ref{st}) becomes
\be (e^{ \pi i
/2})^\ymon (-)^{N_1+\tn_1} \ket{ \{2\};\{1\};\{Y\} } \ .
\ee
Here we
have used the short hand notation $\{i\} = N_i,\tn_i,n_i,w_i$.

At the massless level $\ymom = 0 \Rightarrow i^\ymom = 1$. The
quarter-shift around the circle, therefore, does not play a role in
determining the fate of the moduli; for this reason we will drop the
$\{Y\}$. Similarly zero modes $n_i,w_i$ will be dropped. Note that
the massive states do transform non-trivially under the
quarter-shift. As will be seen when we construct the partition
functions, the theory with the shift and without the shift are very
different.  Moreover, it is worth noting that here we are only
discussing the untwisted sector.  One might be concerned that the
twisted sector includes new massless states.  Indeed, this is often
the case for orbifolds which are not integrated with a shift.
However, as we will see explicitly below, integrating the shift with
the pure orbifold gives masses to all twisted sector
states\footnote{For this to be true one must be careful in the way
the shift orbifold is constructed.  The subtleties in how one
constructs the shift orbifold will be discussed shortly.}.

Fluctuations in the metric/B-field correspond to states
created by acting with the symmetric/antisymmetric
combinations of $\a_{-1}^{\mu} \tilde{\a}_{-1}^{\nu}$.
To see which states are projected out we first
form the eigenstates
\be\label{bosoestates}
\ket{\pm;\{1\},\{2\}} \equiv \ket{\{1\},\{2\}} \pm
\ket{\{2\},\{1\}} \ .
\ee
It is easy to convince oneself that at the
massless level $N_1 + \tn_1 = N_2 + \tn_2 \ \mbox{mod} 2$ and
\eqn{bosoestates} are then eigenstates with eigenvalues
$\pm (-)^{N_1+ \tn_1}$.
Under (\ref{symorb}),
\bea
\ket{\pm;1,1;0,0} &\longrightarrow& \pm \ket{\pm;1,1;0,0}
\label{keepkilla}
\\
\ket{\pm;1,0;0,1} &\longrightarrow& \mp \ket{\pm;1,0;0,1} \ .
\label{keepkillb}
\eea
The orbifold projection requires us to keep only the states with eigenvalue
$+1$. From (\ref{keepkilla}) we see that the fluctuations, $h_{ij}$,
in the torus metric must satisfy $h_{11} = h_{22}$, and from
(\ref{keepkillb}) we see that $h_{12} = 0$.  Fluctuations in the
B-field are not projected out. Since we have quantized around
$ds_{T^2} = \delta_{ij} dX^idX^j$ this is exactly the condition
$\tau = i$.
In the dual (pseudo asymmetric) theory, where $\rho \rightarrow
-1/\rho$, the action of the corresponding orbifold is (\ref{symorb})
along with $X^i_R \rightarrow - X^i_R$.  For states with
$n_i=w_j=0$,
\be \ket{\{1\};\{2\};\{Y\}}
\underbrace{\longrightarrow}_{X^i_R \rightarrow - X^i_R} (-)^{\tn_1
+ \tn_2} \ket{\{1\};\{2\};\{Y\}} \ . \label{tdltyact}
\ee
Noting that, at the
massless level $(-)^{\tn_1 + \tn_2} = -1$, we can see integrating $X^i_R
\rightarrow - X^i_R$ with (\ref{symorb}) introduces an overall
change in sign in (\ref{keepkilla}) and (\ref{keepkillb}).  As
expected, projecting onto states with eigenvalue $+1$ gives
condition $\rho = i$.
\subsubsec{The partition function}
We will now construct the partition function for these orbifolds. First
consider orbifolding the torus by a 90 degree rotation (without the
interpolation). The sector with $a$ twists and $b$
insertions\footnote{This does not hold for $a=b=0$ since we have not
included zero modes.  Its also worth noting that
$\q^a_b \propto 1/Z^{1-a/2}_{1+b/2}$ where $Z^{1-a/2}_{1+b/2}$ is defined in \eqn{joepart}.}
is given by ,
\bea\label{rotorbpart}
&& Z^a_{\ \ b}(\tau) =
\left|\q^a_b\right|^2 \ ; \\
&& \q^a_b \equiv
 \frac{\eta(\tau)}{\vartheta\left[ {\frac{1}{2}
- \frac{a}{4}} \atop  {\frac{1}{2} + \frac{b}{4}} \right](\tau)}
\times \exp\bigl[2\pi i (\f b 4 + \f 1 2) (\f a 4 - \f 1 2)\bigr]
\times (1-e^{\pi i b/2})^{\delta_{a,0}} \ . \label{qab} \eea Because
the two theories are T-dual, the partition function for the $\rho
\rightarrow -1/\rho$ monodrofold should be the same as the $\tau
\rightarrow -1/\tau$ monodrofold.  To check this explicitly one
needs to construct the pure orbifold of the dual theory.  This is a
worthwhile exercise since, from the worldsheet perspective, there is
a priori no reason to expect the two orbifolds to be dual. This
gives further, albeit indirect, evidence that these interpolating
orbifolds are in fact the correct worldsheet description of
monodrofolds.  This is easy to check. $\rho \rightarrow -1/\rho$
corresponds to T-dualizing along both legs of the torus along with a
90 degree rotation.  In terms of our embedding coordinates this is
simply a 90 degree rotation on the left hand side of the string and
a -90 degree rotation on the right hand side of the string. Being
careful with the phases one finds, \bea \label{beforeshifts}
\widetilde{Z}^a_{\ \ b}(\tau) &=&
\q^a_b \bar{\q}^{||-a||}_{||-b||} \\
&=& \left|\q^a_b\right|^2 \label{aftershifts} \eea The ``$||m||$''
in the superscripts and subscripts indicates that one should use the
lowest positive value of ``m''; i.e. m= 0,1,2,3  (for a fourth order
orbifold). Going from (\ref{beforeshifts}) to (\ref{aftershifts}) we
have used known transformation properties of the $\vartheta$
functions. As expected, (\ref{aftershifts}) agrees with
(\ref{rotorbpart}).  Its important that one not over extend the
validity of the expression \eqn{qab}. $\q^a_b\ (\bar{\q}^a_b)$  is
the contribution to the partition function coming from the left
(right) side of the string.  However the derivation only holds for
the symmetric and pseudo asymmetric orbifolds, and one can not
arbitrarily pair a $\q^a_b$ with a $\bar{\q}^c_d$ to form asymmetric
orbifold sectors.  For instance, in the orbifolds discussed in this
subsection there are no terms $\q^2_b \bar{\q}^0_0$.  Such a sector
would seemingly correspond to twisting the left hand side of the
string by 180 degrees while leaving the right hand side of the
string unaffected. As discussed in section \ref{sec:circ}, this is
the incorrect partition trace for such a orbifold, and moreover,
such an orbifold is not modular invariant.

If one were not to integrate the $\tau \rightarrow -1/\tau$
(or $\rho \rightarrow -1/\rho$) orbifold with a shift around the base,
the total partition function would be
\bea Z(\tau) &=&
i V_{10} \int_F {\f {d^2\tau}{16 \pi^2 \alpha^\prime \tau_2^2}}Z^{[10]}(\tau) \ ; \\
Z^{[10]}(\tau) &\equiv&
Z^{[6]}(\tau)\   {\cirpart {R}}^2\  \frac{1}{4} \sum_{a,b} Z^a_{\ \
b}(\tau) \ . \label{symorbpart}
\eea $Z^{[6]}(\tau)$ is the
partition function trace for the fields which do not play a role in
(\ref{wwfmonodromy}) and $\cirpart R$ is the partition function for a
circle of radius $R$; \bea \hspace{-.4in} \cirpart R &=&
\left|\eta(\tau)\right|^{-2} \sum_{n,w = \infty}^{\infty} \exp
\left[ -\pi \tau_2 \left(\frac{\ap n^2}{R^2} + \frac{w^2
R^2}{\ap}\right) + 2\pi i \tau_1 n w \right]  . \label{eqn:cirpart}
\eea

In the partition function for the interpolating orbifold, the $\{a,b\}$
contribution to the shift orbifold must be paired up with $\{a,b\}$
contribution to the pure orbifold.  Namely, \be Z^{[10]}(\tau) =
Z^{[6]}(\tau) \cirpart R \ \frac{1}{4}  \sum_{a,b} Z^a_{\ \ b}(\tau)
\ \cirpart R^a_{\ \ b} \ . \label{intsymtotal} \ee
Where, \bea
\cirpart R &=& \frac{1}{4} \sum_{a,b} \cirpart R^a_{\ \ b} \ .
\label{eqn:cirpartsum}
\eea

As suggested above, to construct the $\cirpart R^a_{\ \ b}$ traces
we will start with a circle of radius $4R$ and orbifold by a
quarter-shift around the circle. Orbifolding by the shift alone
simply quarters the size of circle, $\cirpart {4R} \rightarrow
\cirpart R$.  It would certainly be possible to choose an Nth-order
shift around a circle of radius $NR$.  Indeed, from what we have
already learned, the massless sector coming from the untwisted
sector is unaffected by the shift.  Moreover, as long as both the
shift orbifold and the pure orbifold are modular covariant the
theory will be modular invariant. However, in order to make
connection with the spacetime constructions in \cite{wwf} we argue
that an order $N$ pure orbifold action must be integrated with an
order $N$ shift. To see this we should return to the interpolating
nature of the backgrounds discussed in \cite{wwf}.  There, there are
masses  given to certain fluctuations by introducing nonstandard
boundary conditions.  However,  as one takes the the base to
infinite volume the fields once again become massless. Lets now try
and follow this same story from the worldsheet perspective:

We will start by constructing the partition traces (which we needed to do anyway).
Consider a $N$th order orbifold, namely a
$2\pi R$ shift on a circle of radius $NR$. It is easy to reason out what the
$\cirab a b$ should be: $\cirab 0 0 = \cirNnw n w$
by definition.
The funny notation ``$[n|w]$'' is used
to indicate that the trace sums over integers $n$ and $w$.  For
example $\cirNnw n w$, which is equal to \eqn{eqn:cirpart} after replacing $R$ with $NR$,
contains  a sum over all integer values
of $n$.  However, $\cirNnw {2n} w$ only sums over even 
integers\footnote{This notation is explained further in
appendix \ref{ap:shift}.}, $2n
\in 2 \mathbb{Z}$.
In order to find $\cirab 0 b$ we need to insert
the operator $\exp\left(2 \pi i b n/N \right)$ into the sum
(\ref{eqn:cirpart}). It follows that,
\be \cirab 0 b =
\sum_{q = 0}^{N-1} e^{2\pi i b q/N} \cirNnw {N n + q} w
\ee
In the
$a$-th twisted sector, $\cirab a 0$,
we now have strings which wind $a/N$ of the way around the
original circle. It follows that one must replace  $w$ with $w + a/N$,
\be
\cirab a 0 = \cirNnw n {w + a / N} \ .
\ee
Combining these operations is straightforward.  We find that the sector with $a$ twists and
$b$ insertions is given by the trace
\be\label{shiftab}
\cirab a b =
\sum_{q = 0}^{N-1} e^{2\pi i b q/N} \cirNnw {N n + q} {w + a / N}\ \ .
\ee
As a quick check note that,
\bea
\f 1 N \sum_{a,b} \cirab a b &=&
\f 1 N \sum_{q,a} \underbrace{\biggl[\sum_b e^{2\pi i b q/N}\biggr]}_{=N \delta_{q,0}}
\cirNnw {N n + q} {w + a / N} \nonumber \\
&=& \sum_a \cirNnw {N n} {w + a / N} \nonumber \\
&=& \cirNnw {Nn} {w/ N} \nonumber \\
&=& \mathcal{Z}_R \ \ .
\eea

Having constructed the partition traces it is now easy to see why,
when integrating a shift with an $N$th order pure orbifold, we
require an order $N$ twist.  Suppose for
 $a$-twists $a/N = m \in \mathbb{Z}$. The twisted winding states in the $a$-th twisted sectors
now include a massless mode, since $w\rightarrow w + m \in
\mathbb{Z}$. Taking $R \rightarrow \infty$ the $m=0$ mode does not
drop out of the theory.  It follows that there exist twisted states,
$i.e.$ states which do not exist in the unorbifolded theory, which
do not drop out of the theory as we go to the noncompact limit. Such
an orbifold is not, therefore, ``interpolating'' in the sense
presented above\footnote{We have been a little to slick in this
argument, and in the next section will have too slightly modify our
conclusions.}. Having said this, we have constructed the shift
orbifold for general $N$, and in turn given a world sheet
description to our pure orbifold integrated with an $N$th order
shift.  It would be interesting to try to identify the the
corresponding spacetime physics and contrast it with the monodrofold
discussed above.

Using
(\ref{rotorbpart},\ref{eqn:cirpart},\ref{intsymtotal},\ref{shiftab})
one obtains the full partition function for a $2
\pi/4$ rotation integrated with a quarter-shift around the circle.
Modular covariance of (\ref{rotorbpart},\ref{aftershifts}) follows
trivially from transformation properties of the
$\eta/\vartheta$-functions. The modular covariance of the $\cirab a b$ is
demonstrated in appendix \ref{ap:shift}.  As discussed
in section \ref{sec:circ}, this guarantees the modular invariance of
the total partition function (\ref{intsymtotal}).

\section{Truly asymmetric orbifolds}\label{sec:het}

Lets now turn to the heterotic string. In particular we will focus
on the $SO(32)$ case, though everything can be easily generalized to
the $E_8\times E_8$ case.  The construction in this section involves
an asymmetric orbifold which is truly nongeometric. We'll see that
although from the spacetime perspective this is possible in an
arbitrary number of dimensions, modular invariance forces us to have
either be in $4$ or $8$ dimensions. To do this we will have to go to higher
dimensional fibrations, which is convenient since one is ultimately
interested in getting down to 4-dimensional theories.  To do this we
will make a trivial modification of the the backgrounds discussed in
\cite{wwf}. Rather than looking at $T^2$ fibrations over a $T^2
\times T^{16}$ base we will consider $T^2 \times T^2 \times T^{16}$
fibrations over a $T^2$ base. Here $T^{16}$ is the internal torus of
the heterotic string. \eqn{sugra} is then modified in the obvious
way,
\begin{eqnarray}\label{sugrahet}
\db \rho^i = \db \tau^i = \d \db (\ln \rhot_2 - \ln \rho_2^1 - \ln
\tau_2^1
                        - \ln \rho_2^2 - \ln \tau_2^2 )=0 \ .
\end{eqnarray}
Here $\rho^i,\tau^i$ correspond to the moduli coming from the
$i=1,2$ torus. In addition to \eqn{sugrahet} there is a self-duality
constraint on field strength arising from the internal gauge fields.
As usual we require the gauge fields commute so that the potential
$\sim Tr([A_\mu,A_\nu]^2)$ vanishes, thus the most general Wilson
line background uses only the sixteen $U(1)$ gauge fields in the
Cartan of the heterotic gauge group. Each of these $U(1)$'s can be
associated with a cycle of the internal torus. Requiring that they
are constant along the $T^2$ fiber directions, the self-duality
constraint reduces to the fact that the gauge fields must also be
constant along the base $T^2$ \cite{wwf}, i.e. \be\label{sugrahetA}
\d_{\mu} A_{\nu}^\mathcal{J} = 0 \ . \ee

In particular we will look at the monodromy \be \label{hetmonodromy}
\tau^i \rightarrow -1/\tau^i \ , \ \ \rho^i \rightarrow -1/\rho^i \
, \ \ T^{16} \rightarrow \mathcal{R}_n T^{16}. \ee as we encircle
one of the cycles in the base.  Here $\mathcal{R}_n$ is a reflection
on $n$ of the internal bosons, or in the fermionic picture, $n$ of
the internal complex fermions. As discussed in the beginning of 
section \ref{sec:wwf} this corresponds to a truly nongeometric
background.

\subsubsec{Moduli fixing}
As before, the spacetime boundary conditions fix $\rho^i = \tau^i =
i$.  The boundary  conditions also require $T^{16} \rightarrow
\mathcal{R}_n T^{16}$. Since the gauge fields must be constant
\eqn{sugrahetA} this forces them to vanish on the directions in
which $\mathcal{R}_n$ has a nontrivial action.

From the worldsheet perspective the orbifold projection is quite
similar to that in the previous section. The usual geometric and
B-field moduli in the heterotic string arise from
symmetric/antisymmetric combinations of $\alpha_{-1 \ \mu}
\tilde{\psi}^{NS}_{-1/2 \ \nu}$ acting on the NS vacua. In addition
we have the moduli from heterotic gauge fields which lie in the
Cartan of $SO(32)$ or $E_8\times E_8$ with constant expectation values
along the $T^2$ fibers. In the bosonic formulation these states
arise from $\alpha_{-1}^i\tilde{\alpha}^{\mu}_{-1}$ where $i$ labels
the Cartan $U(1)$ factors.  Orbifolding by a reflection in the $ith$
cycle of the internal torus introduces $(-1)^N_i$ under which
$\alpha_{-1}^i\tilde{\alpha}^{\mu}_{-1}$ is odd. As expected from
the spacetime physics, the states corresponding to the massless
fluctuations of  $A_{\nu}^\mathcal{J}$ are projected out of the
theory.
\subsubsec{The partition function}
In order to construct the world sheet theory we follow the same
steps as in the previous section.  Namely, we first construct the
partition traces for the pure orbifold and show they are modular
covariant.  We then integrate the pure orbifold with a shift around
the circle.

The internal and fermionic contribution to the partition function
for the heterotic string is $Z_{16}Z_\psi^{+*}$\cite{joeii}.
Here $Z_{16}$ is the contribution coming from the internal lattice
and $Z_\psi^{+*}$ is the contribution coming from the fermions.  In
terms of $Z^\alpha_\beta$, as defined in \eqn{joepart}, \bea
&&Z_{16}(\tau) = \frac{1}{2}
  [Z^0_0(\tau)^{16} + Z^0_1(\tau)^{16} + Z^1_0(\tau)^{16} + Z^1_1(\tau)^{16}]
\label{z16} \\
&&Z_\psi^+(\tau) = \frac{1}{2}
  [Z^0_0(\tau)^{4} - Z^0_1(\tau)^{4} - Z^1_0(\tau)^{4} - Z^1_1(\tau)^{4}] \ .
\label{zpsi} \eea Its worth noting that although the product
$Z_{16}Z_\psi^{+*}$ is modular invariant, the individual partition
sums (\ref{z16},\ref{zpsi}) are not\footnote{Under inversions, $\tau
\rightarrow -1/\tau$, the partition sums do transform into
themselves.}. To be precise, under modular shifts $\tau \rightarrow
\tau +1$ \be Z_{16}(\tau + 1) \rightarrow e^{2 \pi i/3} Z_{16}(\tau)
\ , \ \ Z_{\psi}^+(\tau + 1) \rightarrow e^{2 \pi i/3}
Z_{\psi}^+(\tau) \ . \ee This gives us a hint on what to look for as
we proceed.  We will focus on asymmetric reflection orbifolds of the
embedding coordinates and the internal lattice. Since these
orbifolds will leave the fermionic contribution to the partition sum
unaffected, the partition sum coming from orbifolded piece must pick
up a phase under modular shifts.  Keeping this in mind we will relax
the condition of modular covariance, and look for theories which are
modular covariant up to phases.

First consider reflection orbifolds of the internal lattice.  The
details are worked through in appendix \ref{ap:fermi}. Starting
with the $\alpha,\beta$-sector, $Z^{\alpha \ 16}_\beta$, and preforming $a$
twists and $b$ insertions on $n$ of the dimensions one finds 
\be Z^{\alpha\
16}_\beta \longrightarrow (-)^{\alpha(\alpha-1)\beta/2} Z^{\alpha\
16-n}_\beta  Z^{\alpha-a\ n}_{\beta+b}\ . \ee The total partition
function for the internal part is then \be Z_{16/\mathcal{R}} =
\sum_{a,b=0}^3 \lambda^a_b \ , \ee where \be \lambda^a_b \equiv
\sum_{\alpha,\beta = 0}^{1} (\mbox{sgn}_{ab}[\alpha,\beta])^n
Z^{\alpha\ 16-n}_\beta Z^{||\alpha-a|| \ n}_{||\beta + b||} \ \ .
\ee Here $||m|| \equiv (1-(-)^m)/2$, \ie\ $0$ if $m$ is even and $1$
if $m$ is odd.\footnote{Though we are using the same notation
``$||...||$'' as in section \ref{sec:wwf} they have a slightly
different meaning. Here we are restricting ourselves to $m=0,1$
even though we are discussing a 4th order orbifold.} Table \ref{sgn}
gives the values for $\mbox{sgn}_{ab}[\alpha,\beta]$. \small
\TABULAR{l||cccc|cccc|cccc|cccc|}{ \backslashbox{$\alpha,\beta$}{\va
ab}& \vb00 & \vb02 & \vb20 & \vb 22 & \vb01 & \vb03 & \vb21 & \vb 23
& \vb10 & \vb12 & \vb30 & \vb 32 &
\vb11 & \vb13 & \vb31 & \vb 33 \\
\hline \hline 0,0 & + & + & + & + & + & + & - & - & + & - & + & - &
- & + & + & - \\
0,1 & + & + & - & - & + & + & + & + & - & + & + & - &
- & + & - & + \\
1,0 & + & - & + & - & + & - & - & + & + & + & + & + &
+ & + & - & - \\
1,1 & + & - & - & + & - & + & - & + & + & + & - & - &
+ & + & + & + \\
\hline }{\label{sgn}$\mbox{sgn}_{ab}[\alpha,\beta]$:\ \ The
{a,b}-columns have been subdivided into groups which have the same
value for $||a||$ and $||b||$.} \normalsize When orbifolding an even
number of dimensions, $n \in 2\mathbb{Z}$ there are only 4 distinct
sectors, \bea\label{lambdasectors}
\lambda^0_0 &=& 2 Z_{16} \\
\lambda^0_1 &=& Z^{0\ 16-n}_0 Z^{0\ n}_1 +  Z^{0\ 16-n}_1 Z^{0\ n}_0
+
                Z^{1\ 16-n}_0 Z^{1\ n}_1 +  Z^{1\ 16-n}_1 Z^{1\ n}_0 \\
\lambda^1_0 &=& Z^{0\ 16-n}_0 Z^{1\ n}_0 +  Z^{0\ 16-n}_1 Z^{1\ n}_1
+
                Z^{1\ 16-n}_0 Z^{0\ n}_0 +  Z^{1\ 16-n}_1 Z^{0\ n}_1 \\
\lambda^1_1 &=& Z^{0\ 16-n}_0 Z^{1\ n}_1 +  Z^{0\ 16-n}_1 Z^{1\ n}_0
+
                Z^{1\ 16-n}_0 Z^{0\ n}_1 +  Z^{1\ 16-n}_1 Z^{0\ n}_0
\eea Its easy to show that for $n=4m\in 4 \mathbb{Z}$ under $\tau
\rightarrow -1/\tau$ \be \lambda^0_0\rightarrow\lambda^0_0 \ , \ \
\lambda^0_1\rightarrow\lambda^1_0 \ , \ \
\lambda^1_0\rightarrow\lambda^0_1 \ , \ \
\lambda^1_1\rightarrow\lambda^1_1 \ , \label{translambda1} \ee and
under  $\tau \rightarrow \tau + 1$ \be \lambda^0_0\rightarrow
e^{2\pi i/3}\lambda^0_0 \ , \ \ \lambda^0_1\rightarrow e^{2\pi
i/3}\lambda^0_1 \ , \ \ \lambda^1_0\rightarrow e^{2\pi
i/3}(-)^m\lambda^1_1 \  , \ \ \lambda^1_1\rightarrow e^{2\pi
i/3}(-)^m\lambda^1_0 \ . \label{translambda2} \ee Note that each
term picks up a phase $e^{2\pi i/3}$.  This is good news! However,
in the $\lambda^1_b$ sectors there is an additional factor of
$(-)^m$.  If one were not to orbifold any of the worldsheet bosons
it would be necessary to pick $m\in 2\mathbb{Z}_+$ or $n = 8,16$. It
turns out that even after orbifolding the the embedding coordinates
this will be the case.  This is certainly not surprising since we
know that even self dual Euclidean lattices only exist in
$8\mathbb{Z}_+$-dimensions.

Now we would like to orbifold $d$ of the embedding coordinates by a
chiral reflection.  Fortunately this has been worked out for us in
\cite{adp}\footnote{Our $\mathcal{K}^a_b$ corresponds to their
$\mathcal{Z}^{r_L^b}_{r_L^a}$.  Note that the positions of the
twist/insertion super/sub-scripts has been reversed.}. Denote the trace with
$a$ twists and $b$ insertions by $\mathcal{K}^a_b$ and define,
\be\label{bososectors}
{Z_X}^a_b = \bigl( \mathcal{K}^{a}_b \bigr)^d + \bigl( \mathcal{K}^{a+2}_b\bigr)^d +
\bigl( \mathcal{K}^{a}_{b+2} \bigr)^d + \bigl( \mathcal{K}^{a+2}_{b+2} \bigr)^d \ .
\ee
It is the ${Z_X}^a_b$ which multiply the sectors in
\eqn{lambdasectors}. As in \cite{adp} we will express these in terms
of $\zeta$-functions.  There is a brief review of these functions in
appendix \ref{ap:zeta} and a more detailed review in the appendices
of \cite{adp}. 
\begin{eqnarray}
{Z_X}^0_0 &=& ( |\zeta_{00}|^{2} + |\zeta_{\frac{1}{2} 0}|^2 )^d
+ ( \zeta_{00}\bar{\zeta}_{\frac{1}{2} 0} + \zeta_{\frac{1}{2} 0}\bar{\zeta}_{0 0})^d \nonumber \\
&&
+ ( |\zeta_{00}|^{2} - |\zeta_{\frac{1}{2} 0}|^2 )^d
+ ( \zeta_{00}\bar{\zeta}_{\frac{1}{2} 0} - \zeta_{\frac{1}{2} 0}\bar{\zeta}_{0 0})^d
\label{zxoo}\\
\frac{1}{2} {Z_X}^0_1 &=& \zeta_{0\frac{1}{2}}^d \bar{\zeta}_{00}^d +
                e^{i \pi d/4} \zeta_{0\frac{1}{2}}^d \bar{\zeta}_{\frac{1}{2}0}^d \\
\frac{1}{2} {Z_X}^1_0 &=& \zeta_{\frac{1}{4}0}^d \biggl[
(\bar{\zeta}_{00}+\bar{\zeta}_{\frac{1}{2}0})^d+
                e^{i \pi d/4}(\bar{\zeta}_{00}-\bar{\zeta}_{\frac{1}{2}0})^d \biggr] \\
\frac{1}{2} {Z_X}^1_1 &=& e^{i \pi d/8}\zeta_{\frac{1}{4}\frac{1}{2}}^d
\biggl[ (\bar{\zeta}_{00}
                + e^{-i \pi /2}\bar{\zeta}_{\frac{1}{2}0})^d+
                e^{i \pi d/4}(\bar{\zeta}_{00}- e^{-i \pi /2}\bar{\zeta}_{\frac{1}{2}0})^d \biggr]
\end{eqnarray}
Using the $\zeta$-function transformation properties it is easy
to show if $d=4$, under $\tau \rightarrow -1/\tau$ \be
{Z_X}^0_0\rightarrow{Z_X}^0_0 \ , \ \ {Z_X}^0_1\rightarrow{Z_X}^1_0
\ , \ \ {Z_X}^1_0\rightarrow{Z_X}^0_1 \ , \ \
{Z_X}^1_1\rightarrow{Z_X}^1_1 \ , \label{transx1} \ee and that under
$\tau \rightarrow \tau + 1$ \be {Z_X}^0_0\rightarrow {Z_X}^0_0 \ , \
\ {Z_X}^0_1\rightarrow {Z_X}^0_1 \ , \ \ {Z_X}^1_0\rightarrow
{Z_X}^1_1 \  , \ \ {Z_X}^1_1\rightarrow {Z_X}^1_0 \ .
\label{transx2} \ee 
Using (\ref{translambda1},\ref{translambda2},\ref{transx1},\ref{transx2}), and defining 
$Z^{[10-d]}$ to be the partition trace coming from the unorbifolded bosons,
we find the total partition function, 
\be 
{Z} =
i V_{10} \int_F {\f {d^2\tau}{16 \pi^2 \alpha^\prime \tau_2^2}} Z^{[10-d]}
Z_\psi^{+*}(\lambda^0_0 {Z_X}^0_0 +\lambda^0_1 {Z_X}^0_1 +
\lambda^1_0 {Z_X}^1_0 +\lambda^1_1 {Z_X}^1_1) \ ,
\ee 
is modular covariant for $n=0,8,16$ $d=0,4,8$.

We can now integrate this orbifold with a shift to build the
partition function corresponding to the monodrofold
\eqn{hetmonodromy}. For truly asymmetric orbifolds there is a new subtlety 
which arises in choosing
the order of the shift orbifold.  Recall that we have argued that an
order $N$ action should be integrated with an order $N$ shift. Our
reasoning came from the fact that we wanted to make sure that in the
$R \rightarrow \infty$ limit the twisted sectors dropped out of the
theory. The orbifold currently being discussed at first sight appears
to be 4th order. However, the spacetime monodromy appears to be
second order, and moreover our naive concept of a reflection is 
second order.  Recall that, as discussed in section \ref{sec:circ}, the 4th
order nature of the chiral reflection is due to the phase ambiguity which 
arises when splitting up the rotation $R(\theta)$ up into a left/right-handed
piece; $R(\theta)=R_L(\theta)R_R(\theta)$ \cite{adp}.
With this in mind let's take another look at the order
of this orbifold. For each piece of the partition trace in \eqn{z16}
the action is 4th order. However, when orbifolding you must act on
all 4 sectors in \eqn{z16}. Summing these we see that (in
$d=4\mathbb{Z}$) the operation is actually second order. The
asymmetric orbifold on the spacetime bosons is trickier. As derived,
the sum in \eqn{bososectors} was not forced on us at the outset, but
rather a grouping due to the second order nature of the internal
lattice reflection. Moreover, the individual elements,
$\mathcal{K}^{a}_b$, are not themselves modular covariant. If we
then try and integrate them with a 4th order shift we will not have
a modular invariant theory.  However, if we integrate with a 2nd
order shift the terms again group together properly and the theory
is modular invariant. The question we must ask is then, does the
orbifold integrated with the 2nd order shift correspond to the
spacetime monodrofold defined by \eqn{hetmonodromy}?  This question
is equivalent to asking if one is allowed to keep sectors, in the
$R\rightarrow \infty$ limit, in which there has been a chiral
rotation of 360 degrees. Its our belief that since from a purely
spacetime perspective this action is trivial we can still identify
this orbifold with the monodrofold given in \eqn{hetmonodromy}. Note
that this is very different than the cases discussed in the previous
section where the order 2 twist has definite spacetime implications.
We will return to this issue in the discussion section.

The total partition function for the interpolating
orbifold corresponding to the spacetime defined by the monodromy \eqn{hetmonodromy} is then
\bea {Z} =
i V_{10} \int_F {\f {d^2\tau}{16 \pi^2 \alpha^\prime \tau_2^2}}Z^{[10-d]} \mathcal{Z}_{R}
Z_\psi^{+*} \sum_{a,b = 0}^1 \lambda^a_b {Z_X}^a_b {\mathcal{Z}_R}^a_{\ \ b} \ .
\eea
Here ${\mathcal{Z}_{R}}^a_{\ \ b}$ is as defined in \eqn{shiftab} for a
second order shift, $d$ (the number of embedding coordinate which are orbifolded) is $0,4$ or $8$.
and the number of reflections on the internal lattice is $0,8$ or $16$.

%
%
%
%

\section{Monodrofolds involving modular shifts}\label{sec:shift}

So far in this paper we have concentrated on monodromies in which
$\rho \rightarrow -1/\rho$ and/or $\tau \rightarrow -1/\tau$.
As a final example we would like to consider monodromies which include
modular shifts, $\tau \rightarrow \tau + 1 ~ (~\rho \rightarrow \rho
+ 1~)$.  A shift alone does not admit constant moduli as solutions
and is therefore not in the class of backgrounds we are
discussing\footnote{In the discussion section we will say more about
monodromies which force the moduli to vary.}.  We can however find
monodromies with constant moduli solutions if we combine modular
inversions with modular shifts. There is some subtlety to combining
these actions.  For example, there are two seemingly different
monodromies arising from combining a single inversion ($R:~\tau
\rightarrow -1/\tau$) and a single shift ($S:~\tau \rightarrow \tau
+ 1$): \be\label{srrs} RS:~\tau \rightarrow - \f 1 {\tau + 1} \ , \
\ SR:~\tau \rightarrow - \f 1 \tau + 1 \ . \ee However, let's be
careful before proceeding. The constant solutions for these
monodromies are, $\tau^{RS}_\pm \equiv -1 \pm i \sqrt{3}$ and
$\tau^{SR}_\pm \equiv 1 \pm i \sqrt{3}$ respectively.  Clearly
$\tau^{RS}_\pm =-\tau^{SR}_\mp$, and therefore the constant modulus
describes the same torus  (in
different fundamental domains). From the spacetime picture we see
that the massless fluctuations around our background are the same
for the $RS$ theory and the $SR$ theory. From the worldsheet we will
be able to make a much stronger statement, namely any differences
between the $RS$ theory and the $SR$ theory do not contribute to
the partition trace. This is reflective of the fact that when one
says an orbifold is abelian they mean that the point group (the
group coming just from the rotation generators) is abelian.  This is
trivially true in these orbifolds since $S$, as will be discussed
below, is a pure shift.

In constructing the orbifold in these two theories the operation
corresponding to $R$ is of course the same rotation by 90-degrees
which we discussed in section \ref{sec:wwf}.  $S$, which takes $\tau
\rightarrow \tau +1$, is simply $\exp[2\pi R (p_1 + p_2)]$, where
$p^i$ is the momentum along the coordinates of the fiber, $X^i$.  To
make this clear, think of the torus fiber as a circle fibered over another
circle.  Orbifolding by a simultaneous shift in the fiber circle and the base circle 
gives a theory such that, upon translation
around the base the circle, the fiber must shift by $2\pi R$.
Another way to think about this is if you start with as square torus
and orbifold along the diagonal this gives a skewed torus with
$\tau_2$ equal to the length of the diagonal (divided by $2\pi$).

In order to insure that the partition function is modular invariant,
we will once again demand that the orbifold without the shift is
modular covariant. Here we must be careful in defining what we
mean by modular covariance. Defining \be \qq \nonumber \ee to be the
sector with $a$ $S$-twists, $b$ $S$-insertions,
$\alpha$ $R$-twists, and $\beta$ $R$-insertions, the correct modular covariance
requirement is, \bea \qq {\longrightarrow}\ \Q a \alpha {b-a}
{\beta-\alpha} \eea under $\tau \rightarrow \tau + 1$, and \bea \qq
{\longrightarrow}\ \Q b \beta {-a} {-\alpha} \eea under $\tau
\rightarrow -1/\tau$.

It is trivial to construct our $Q[...]$'s using the traces given in
\eqn{rotorbpart} and \eqn{shiftab}.  By definition,
\bea
\label{modshift}
\Q a 0 b 0 &=& {\mathcal{Z}_{R_1}}^a_{\ \ b}~{\mathcal{Z}_{R_2}}^a_{\ \ b}  \\
\Q 0 \alpha 0 \beta &=& |\mathcal{Q}^\alpha_\beta|^2\ . \eea
To construct the other sectors
consider first acting with elements of $R$.  For the twisted sectors
there are no zero modes, and for the untwisted sectors with
insertions the zero modes drop out of the trace.  It follows in
neither case does the orbifold by the shift contribute to the
partition trace: \bea \Q a \alpha b \beta &=&
|\mathcal{Q}^\alpha_\beta|^2  \ \ \ \ \ \ [{\rm Except\ for}\
\alpha=\beta=0 ] \eea If one were first to act with $S$ the story
would be similar. The operators in the untwisted sectors would have different eigenvalues
(since you'd be acting with $p^i$'s before the rotation) but since
the only contribution to the partition trace comes from $p^i=0$ this
does not effect our final expression.

So far we have focused on a single shift and a single inversion.
However the results are easy to generalize.  Since $R^{2n}\propto 1$ orbifolds
with an even number of 90-degree rotations will not admit constant $\tau$ solutions.  Moreover,
$R^{2m + 1} = (-)^m R$.  Replacing a single $S$ with $S^N$ simply involves replacing
the radii $R_i$ in \eqn{modshift} with $N R_i$.  Since the orbifold actions
corresponding to $S$ and $R$ are abelian (in the sense discussed above) $R^{2m+1} S^N$ is
the most general monodromy admitting constant $\tau$ solutions.
 
We can now integrate this orbifold with shift in the
straightforward way.  As before we want  all twisted sectors to be massive
and therefore the order of the shift must be the sum of the order of 
$R^{2m + 1}$ and $S^N$ orbifolds, $i.e.$ 5. The total partition function for 
the orbifold  corresponding to space times in which a $T^2$ fiber undergoes ${2m + 1}$ modular 
inversions and $N$ modular shifts is then
\bea 
{Z} =
(-)^m i V_{10} \int_F {\f {d^2\tau}{16 \pi^2 \alpha^\prime \tau_2^2}}Z^{[6]}(\tau)
\sum_{a,b,\alpha,\beta} \qqt_N
{\mathcal{Z}_R}(\tau)^{a +\alpha}_{\ \ b+\beta} \ . 
\eea
Where 
\bea
\qqt_N =
\left\{\begin{array}{ll}
\cirpart {NR_1}^a_{\ \ b}~\cirpart {NR_2}^a_{\ \ b} & \mbox{\ \ \ \ \ if $\alpha=\beta=0$} \\
|\mathcal{Q}(\tau)^\alpha_\beta|^2  & \mbox{\ \ \ \ \ otherwise}
\end{array}
\right.
\ \ \  .
\eea
To construct the pseudo asymmetric dual the $S$ insertions will involve winding
rather than momentum, and the $R$ insertions are exactly as discussed in section \ref{sec:circ}.

%
%
%
%

\section{Discussion}

A crucial part of the development of string theory lies in identifying, and 
exploring the nature of, elements which are not present in the classical
point particle theory.
Nongeometric backgrounds in particular serve as important clues to how we might move
away from the classical formulations of geometry on which string theory is 
so dependent.  
It has been shown from the spacetime perspective
that by allowing various moduli to undergo monodromies in the full stringy duality group
one may construct a broad class of nongeometric 
backgrounds~\cite{dh,wwf,hmw,umanifolds}.
That such backgrounds can be built by incorporating stringy symmetries with a somewhat conventional
spacetime approach may be surprising, and one might
expect that this approach renders certain aspects
difficult to discern.  Doubled formalisms \cite{hull,bem1}, as well as the work being done with
 $SU(3)$ structures \cite{su31,su32,su33} and generalized complex geometry \cite{gcg,marco} are perhaps 
the first steps towards shedding light into some of these areas.

Another obvious approach (and the one taken in this paper)
is through the worldsheet. In this paper, we have used orbifold methods to 
construct nongeometric backgrounds, and argued that they
correspond to the spacetimes discussed in \cite{dh,wwf}.
More precisely, we make explicit through several examples
the connection between interpolating orbifolds
and spacetime duality twists.
By constructing one-loop partition functions and checking modular invariance, it has
become clear that generic nongeometric backgrounds arising from duality twists
will not have simple orbifold constructions.   
On the other hand, we have been able to find several positive results. 
We have constructed 
what we called pseudo nongeometric backgrounds.  These backgrounds
are dual to geometric backgrounds and as a result inherit the
obvious consistency in the dual formulation.  Additionally we have 
been able to formulate backgrounds which are truly nongeometric in
the sense of not being dual (within the context of the perturbative stringy duality group) 
to any geometric solution.

In no sense has this work been exhaustive.
We have limited ourselves to interpolating orbifolds with relatively little
inclusion of shift symmetries and/or non-abelian point groups. By allowing for
a greater variety in the orbifold groups one should be able to construct
many more consistent examples.  
At present there seems little one can say on the pattern
of consistent nongeometric backgrounds in full.  
A key open problem is to classify on general grounds, as opposed to the
example by example approach taken here,
which duality twists lead to modular invariant theories.

All of the above constructions involve spacetime monodromies with fixed points.  
Requiring constant moduli solutions was necessary to insure
that our base would be a torus.  There is even stronger motivation for this 
restriction since if we allow the moduli to vary in these simple models
we would no longer have a solution to the string equations of motion.  
This being said, in more complicated 
compactifications with additional fluxes turned on it should be possible
to find spacetimes which exhibit monodromies without fixed points.
In fact these are exactly the types of backgrounds discussed in \cite{stw}.
It is reasonable to ask if there are orbifold constructions of these models.
Consider a simple example of a
``twisted torus'' \cite{twstdcrc}, where \be\label{twstT} ds^2 = dx^2 +
dy^2 + (dz + Mx~dy)^2 \ \ee and the B-field vanishes. 
Globally we have the identifications $(x,y,z)\simeq
(x,y+1,z)\simeq (x,y,z+1)$ and \be\label{funnyid} (x,y,z) \simeq
(x+1,y,z-My) \ . \ee The $y$-dependent shift in \eqn{funnyid} is
needed to make the torus metric is single valued. 
This describes
a torus fibered over a circle base (whose base is in the $x$ direction) with 
complex structure $\tau = Mx + i$ and kahler modulus $\rho = i$. 
As the fiber traverses the base $x \rightarrow x +1$ and subsequently $\tau \rightarrow
\tau + M$.  The identification \eqn{funnyid} is needed to glue the torus 
with complex structure $\tau+1$ back to the the torus with complex structure $\tau$.
Let's now try to impose the boundary conditions \eqn{funnyid} through
orbifolding.  Blindly following the methods developed in this paper
the partition function would contain pieces of the form
\be\label{nofix}
\sum_{a,b} \cirpartB {R_x} ^a_{\ \ b} ~\cirpartB {R(y)_z} ^a_{\ \ b} \ . 
\ee 
Here
$\cirpartB {R_x} ^a_{\ \ b}$ are again the shift orbifold partition
traces \eqn{shiftab}. $\cirpartB {R(y)_z} ^a_{\ \ b}$ should be
something similar, however we should be careful. We want to shift
$z$ by a distance $y$.  If $y$ was equal to some rational number we
could in principle always do this (although it could involve
orbifolds whose order approached infinity).  However, $y$ is a continuous
variable and for irrational $y$ this can not even be done in principle since it
would necessarily involve an infinite order orbifold.  Moreover, it is unclear how
to properly combine \eqn{nofix} with the partition 
function coming from the
$y$ embedding coordinate.
Compare this to
integrating a rotation: there the rotation operator which is being
integrated with a shift is not dependent on any of the coordinates
of the fiber or base and all of these complications are avoided.  
It is not surprising that this is difficult.
Indeed, were we able to form this orbifold theory, with a position
dependent orbifold action, we would be able to describe strings
propagating on curved backgrounds!

Before concluding we would like to return to an observation made in
section \ref{sec:het}.  To review: There is a phase ambiguity which 
arises when splitting up a rotation into a left and right handed
chiral rotations.  As a result, the chiral reflection, for example, 
is 4th order rather than 2nd order.  Moreover, using the results of
\cite{adp} we see that the partition traces which arise are not
modular covariant.  However, if one properly groups these partition
traces (as in \eqn{bososectors}) one obtains sectors which appear to 
arise from a modular covariant 2nd order theory (\ref{zxoo}-\ref{transx2}).  
This is suggestive and leads us to make the following proposal.  
When constructing truly asymmetric
orbifolds one should should always sum over spacetime equivalent sectors in order
to get a modular covariant theory.
In the chiral reflection example this means that the untwisted sector 
without insertions should include the 4 separate partition traces which
arise from twisting or inserting with the identity or a 360 degree chiral shift. 
As shown above, this suggestion certainly works and has good physical
motivation for the chiral reflection.  However, we have only worked 
through a single example.  One still needs to ask how generic a feature it is
that truly asymmetric orbifolds, which are modular invariant but not modular covariant, 
maybe resummed into
lower order modular covariant partition traces.  Perhaps more importantly,
we should ask if there is a way of developing a formalism which uses this 
perspective to avoid the difficulties which arise in the standard approach
to asymmetric orbifolds. The authors of this paper a currently investigating
these issues.

%
%
%
%
%
%

\appendix

\section{Modular covariance of the shift orbifold}\label{ap:shift}

Since modular covariance is a stronger condition than modular
invariance it is worthwhile check explicitly that the elements of
the shift orbifold \eqn{shiftab} are indeed modular covariant. In
addition to checking modular covariance, this section offers a more
explicit description of these partition function elements.

First we will define,
\be
\Omega[\alpha, \beta;\gamma, \eta] \equiv
\sum_{n,w} \exp\Bigl[-\pi \itau \bigl( (\alpha n + \beta)^2 +
(\gamma w + \eta)^2 \bigr)
          + 2 \pi i \rtau (\alpha n + \beta)(\gamma w + \eta)\Bigr] \  .
\ee Note that $\Omega[1,0;1,0]$ is proportional to the circle
partition function (\ref{eqn:cirpart}) at self dual radius,
$\mathcal{Z}_{\circ}$. Making contact with our earlier notation, \be
\Omega[\alpha, \beta;\gamma, \eta] =
|\eta(\tau)|^2\mathcal{Z}_{\circ}[{\alpha n + \beta}|{\gamma w
+\eta}] \ee Taking $[\alpha, \beta;\gamma, \eta] = [N,q;1,a/N]$ as
in \eqn{shiftab} it is trivial to show that ${\mathcal{Z}_{\f \circ
N}}^a_{\ \ b}$ is properly behaved under modular shifts, $\tau
\rightarrow \tau + 1$.  Moving on to modular inversions, $\tau
\rightarrow -1/\tau$, its straight forward to show $\Omega[\alpha,
\beta;\gamma, \eta]$ becomes $\widetilde{\Omega}[\alpha,
\beta;\gamma, \eta]$, where, \bea \label{omegatilde}
\widetilde{\Omega}[\alpha, \beta;\gamma, \eta] \equiv |\tau|
\frac{e^{-2\pi i \eta \beta/\rtau}}{\alpha \gamma} \sum_{p,m}
\exp\Bigl[-\pi \itau \bigl(\frac{p^2}{\gamma^2} +
\frac{m^2}{\alpha^2} \bigr)
 + 2 \pi i \rtau (\frac{m}{\alpha} + \frac{\eta}{\rtau})
                          (\frac{p}{\gamma} + \frac{\beta}{\rtau})\Bigr]
\eea
Now consider the partition trace
$\mathcal{Z}_{\circ}[{N n + \alpha}|{w+a/N}] = \Omega[N,\alpha;1,a/N]/|\eta(\tau)|^2$.
Using \eqn{omegatilde} we find that under $\tau \rightarrow -1/\tau$
this becomes,
\bea\label{alexsmells}
|\tau| \sum_{p,m}
\biggl[\f {e^{2 \pi i m \alpha/N}} N \biggr] e^{2\pi i a p / N}
\exp{\Bigl[-\pi \tau_2 (p^2 + \f {m^2}{N^2}) + 2\pi i \tau_1 p (\f m N)\Bigr]}
\eea
It then follows from (\ref{shiftab}) and (\ref{alexsmells}) that,
\bea
{\mathcal{Z}_{\f \circ N}} (-1/\tau)^a_{\ \ b} &=&
\sum_{p,m}
\mathbf{P[b]}~e^{2\pi i a p / N}
\exp{\Bigl[-\pi \tau_2 (p^2 + \f {m^2}{N^2}) + 2\pi i \tau_1 p (\f m N)\Bigr]} \\
&=& \sum_{p,\tilde{m}}
e^{2\pi i a p / N}
\exp{\Bigl[-\pi \tau_2 (p^2 + (\tilde{m} - \f b N)^2)
+ 2\pi i \tau_1 p (\tilde{m} + \f b N)\Bigr]} \\
&=& {\mathcal{Z}_{\f \circ N}}(\tau)^{-b}_{\ \ a}
\eea
Here we have used the fact that
\be
\mathbf{P[b]} \equiv
\sum_{\alpha = 0}^{N-1} \biggl[\f {e^{2 \pi i \alpha (m+1)/N}} N \biggr]
\ee
is a projector taking $m\rightarrow \tilde{m} = 4m-b$.  Here we have worked
at the self-dual radius to avoid clutter but by rescaling $\alpha$ and $\gamma$ it is
trivial to extend this proof for all radii.

\section{Fermionic partition traces}\label{ap:fermi}

In this appendix we will discuss the contribution to orbifold
partition traces coming from fermions.  Since this is a key part of
the construction in section \ref{sec:het} we felt that it was
important to discuss some of the subtleties which arise.  Ultimately
we will be interested in orbifolds of the heterotic string. We will
however simultaneously discuss the type II orbifolds due to there
similarities.  Namely, in the type II case the fermionic partition
trace takes the form \cite{joeii} \be Z_\psi^\pm = \frac{1}{2} \bigl[
Z^0_0(\tau)^4 -  Z^0_1(\tau)^4
                              - Z^1_0(\tau)^4 \mp Z^1_1(\tau)^4 \bigr]\ ,
\label{typeiinoorb} \ee and in the heterotic string the internal
directions lead to the partition trace,
 \be
Z_{16} = \frac{1}{2} \bigl[ Z^0_0(\tau)^{16} + Z^0_1(\tau)^{16}
                         +Z^1_0(\tau)^{16} + Z^1_1(\tau)^{16} \bigr]\ ,
\ee The exact definition of $Z^\alpha_\beta$ is given below
(\ref{joepart}). The important point for now is to note the
similarities between the traces. It follows that going from the
heterotic orbifold to the type II orbifold simply requires changing
some signs and powers.

Consider fermionic orbifolds under the action,
\be
\label{fermiorb} \psi(z)^a \rightarrow e^{2 \pi i \lambda} \psi(a)^a
\ .
\ee
Here $\psi^a$ is a complex fermion and $\lambda \in
\mathbb{Z}/2$. The action appears at first glance to be second
order, but we must be careful since we are dealing with fermions.
In particular, twisting by (\ref{fermiorb}) twice generates a
spectral flow.  Namely the ground state before the double twist
becomes and excited state in the doubly-twisted theory.

After a $\lambda$ twist\footnote{Note that taking $\lambda = 0$
gives the untwisted trace as well.} the fermionic part of the stress
energy tensor is, \bea L_0^\psi &=& \frac{1}{2} \sum_{n \in
\mathbb{Z}} (n + \lambda)
:\bar{\psi}^a_{-n-\lambda} \psi^a_{n+\lambda}: \\
&=& \frac{1}{2} \sum_{n \geqslant 0} (n + \lambda)
\bar{\psi}^a_{-n-\lambda} \psi^a_{n+\lambda} -  \frac{1}{2} \sum_{n
< 0} (n + \lambda)
\psi^a_{n+\lambda} \bar{\psi}^a_{-n-\lambda} + a^\psi \\
&=& \frac{1}{2} \sum_{n \geqslant 0} (n + \lambda)
\bar{\psi}^a_{-n-\lambda} \psi^a_{n+\lambda} +  \frac{1}{2} \sum_{n
\geqslant 0} (n + 1 - \lambda)
\psi^a_{-n - 1 + \lambda} \bar{\psi}^a_{n + 1-\lambda} + a^\psi \\
&\equiv& \bar{l}_0 + l_0 + a^\psi \label{set} \eea Here $a^\psi$ is
the zero p.t. energy. The groundstates of the sector twisted by
$\lambda$ is given by, \be \psi_\lambda \ket{0}_\lambda =
\bar{\psi}_{1-\lambda} \ket{0}_{\lambda} = 0 \ \ . \ee

Its important to ask how the reflection insertions act on these
ground states.  Define $\mathcal{R}$ to be a 180 degree rotation in
the untwisted theory.  It follows that \bea && \ket{0}_0 \equiv
\ket{+} \Rightarrow \mathcal{R}^\beta \ket{+} = i^\beta \ket{+}
\label{rgs} \\
&& \ket{0}_{1/2} \equiv \ket{0;\mns}
\Rightarrow \mathcal{R}^\beta \ket{0;\mns} = \ket{0;\mns} \\
&& \ket{0}_1 \equiv \ket{-} = \bar{\psi}_{0} \ket{+} \Rightarrow
\mathcal{R}^\beta \ket{-} = i^{-\beta} \ket{-}
\label{trgs}\\
&& \ket{0}_{3/2} = \bar{\psi}_{-1/2} \ket{0;\mns}
\Rightarrow \mathcal{R}^\beta \bar{\psi}_{-1/2} \ket{0;\mns} \nonumber \\
&& \ \ \ \ \ \ = (-)^\beta \bar{\psi}_{-1/2} \mathcal{R}^\beta
\ket{0;\mns} = (-)^\beta \bar{\psi}_{-1/2} \ket{0;\mns} \eea
Defining $\alpha = 1 - 2 \lambda$, we can see \be \mathcal{R}^\beta
\ket{0}_{\alpha} = e^{\pi i \alpha \beta / 2} \ket{0}_\alpha \ee

Writing down the sum over all $\lambda$-twisted states,
\be
\ket{\lambda} \equiv \prod_{m=1} \ \sum_{F_m,\bar{F}_m = 0}^1
\psi_{-m+\lambda}^{b \ F_m} \bar{\psi}_{-m-\lambda+1}^{b \
\bar{F}_m} \ket{0}_\lambda \ ,
\ee
(here $b$ is being summed from 1
to 4) and noting that,
\bea
&& l_0 \ket{\lambda} = \prod_{m=1} \
\sum_{F_m,\bar{F}_m = 0}^1 (m-\lambda) F_m \psi_{-m+\lambda}^{b \
F_m} \bar{\psi}_{-m-\lambda+1}^{b \ \bar{F}_m}
\ket{0}_\lambda \\
&& \bar{l}_0 \ket{\lambda} = \prod_{m=1} \ \sum_{F_m,\bar{F}_m =
0}^1 (m+\lambda-1) \bar{F}_m \psi_{-m+\lambda}^{b \ F_m}
\bar{\psi}_{-m-\lambda+1}^{b \ \bar{F}_m}
\ket{0}_\lambda \\
&& \mathcal{R}^\beta \ket{\lambda} = \prod_{m=1} \
\sum_{F_m,\bar{F}_m = 0}^1 (-)^{\beta (F_m+\bar{F}_m)} e^{\pi i
\alpha \beta /2} \psi_{-m+\lambda}^{b \ F_m}
\bar{\psi}_{-m-\lambda+1}^{b \ \bar{F}_m} \ket{0}_\lambda \ ,
\label{lastz}
\eea
it is clear that,
\bea
\mbox{Tr}_\alpha[q^{L_0^\psi} \mathcal{R}^\beta] \equiv
Z^\alpha_\beta &=&
q^{(3\alpha^2-1)/24} e^{\pi i \alpha\beta/2} \nonumber \\
&\ &\times \prod_{m=1}^{\infty}\bigl[ 1 + e^{\pi i
\beta}q^{m-(1-\alpha)/2}\bigr] \bigl[ 1 + e^{-\pi i
\beta}q^{m-(1+\alpha)/2}\bigr]
\label{joepart}\\
&=&\frac{1}{\eta(\tau)}\vt \alpha \beta (\tau)
\label{trans1} \ .
\eea
Its worth noting that since we are twisting by
$\alpha = 1 - 2\lambda$,
but we are inserting $\mathcal{R}^\beta$, rather than
$\mathcal{R}^{1-2\beta}$, we have effectively redefined
what we mean by an untwisted fermion.  In particular,
$\alpha = 0$ corresponds to the untwisted $NS$ sector.

These are not, however, valid partition traces for the $\alpha=2,3$
twisted sectors! In deriving (\ref{joepart}) we inserted the same
rotation operator into each sector. Just as the operators $\psi_r$'s
(and consequently $L_0$ and so forth), themselves get twisted, any
operator you insert in a twisted sector must also be twisted. We'll
explicitly construct this operator below, but first lets try to
clarify the main idea. Just as the groundstates depend on
``$\lambda$'' so does the reflection operator, $\mathcal{R}
\rightarrow \mathcal{R}_\lambda$.  Twisting twice takes
$\ket{0}_\lambda,\mathcal{R}_\lambda \rightarrow
\ket{0}_{\lambda+1}, \mathcal{R}_{\lambda+1}$, where action of
$\mathcal{R}_{\lambda+1}$ on $\ket{0}_{\lambda+1}$ is identical to
the action of $\mathcal{R}_{\lambda}$ on $\ket{0}_{\lambda}$.  For
example \be \mathcal{R}_{1/2} \ket{0}_{1/2 + 1} =
\mathcal{R}_{1/2}\bar{\psi}_{-1/2}\ket{0}_{1/2} = - \ket{0}_{1/2 +
1} \ee However, when putting a $\mathcal{R}_\lambda$-insertion in
our partition function we need to use twisted operator,
$\mathcal{R}_{1/2+1}$, \be \mathcal{R}_{1/2 + 1} \ket{0}_{1/2 + 1} =
 + \ket{0}_{1/2 + 1}
\ee

We will now construct $\mathcal{R}_\lambda$ explicitly and check
that it does indeed behave in this manner.  $\mathcal{R}_\lambda$
can easily be expressed in terms of fermion number, however, there
is one subtlety. As mentioned above, we are in essence discussing
spectral flow (which includes intermediate values of $\lambda$).  It
follows that the fermion number operator we write down {\it must} be
able to smoothly vary with $\lambda$. For this reason we choose to
work with the following fermion number operator, \be
\mathcal{F}_{\lambda} = \sum_{r \in Z + \lambda} :\bar{\psi}_{-r}^a
\psi_r^a: = \sum_{n \geqslant 0} \bar{\psi}_{-n-\lambda}^a \psi_{n +
\lambda}^a - \sum_{n \geqslant 0} \psi_{-n-1+\lambda}^a
\bar{\psi}_{n+1-\lambda}^a \ . \ee Its easy to check that the
rotation operator, \be \mathcal{R}_\lambda = e^{\pi i
\mathcal{F}_{\lambda}} \ \ , \ee gives a rotation by $\pi$ when
acting on worldsheet fermions.

Now consider taking $\lambda \rightarrow \lambda + 1$.  The new
ground state, $\ket{0}_{\lambda +1}$, satisfies, \be
\psi_{n+\lambda+1} \ket{0}_{\lambda + 1} = \bar{\psi}_{n-\lambda}
\ket{0}_{\lambda + 1} = 0 \ \ ; \ \ \ n \geqslant 0 \ \ , \ee so
that the operator, \be \mathcal{F}_{\lambda+1}= \sum_{n \geqslant 0}
\bar{\psi}_{-n-1-\lambda}^a \psi_{n + 1+\lambda}^a - \sum_{n
\geqslant 0} \psi_{-n+\lambda}^a \bar{\psi}_{n-\lambda}^a  \ \ , \ee
is still normal ordered (with respect to the new vacuum) and
subsequently behaves exactly the same way as $\mathcal{F}_{\lambda}$
behaved on $\ket{0}_\lambda$.  Note however the operator
$\mathcal{F}_{\lambda}$ is not normal ordered with respect to the
new vacuum since the, $\bar{\psi}_{-\lambda} \psi_{\lambda}$ term is
not normal ordered.  Commuting these operators we can see that
$\mathcal{F}_{\lambda} = \mathcal{F}_{\lambda+1} + 1$ or that \be
\mathcal{R}_{\lambda+1} = e^{\pi i} \mathcal{R}_\lambda \ee

Lets now return to our expression (\ref{joepart}).  It should now be
clear that for the $\alpha = 0,1$ sectors \eqn{joepart} is correct, this is
correct but for the $\alpha = 2,3$ sectors the terms with $\beta =
1,3$ are off by a minus sign. The correct $\alpha,\beta$ sectors for
our orbifold are then, \be \mathcal{P}^\alpha_\beta = (-)^{\alpha
(\alpha - 1) \beta / 2} Z^\alpha_\beta \ee Note that this has the
properties, \be \mathcal{P}^{\alpha+2}_\beta = (-)^\beta
\mathcal{P}^\alpha_\beta \ \ , \ \ \ \mathcal{P}^\alpha_{\beta+2} =
(-)^\alpha \mathcal{P}^\alpha_\beta \label{trans2} \ee It follows
that if we start with $\alpha,\beta$-sector, $Z^{\alpha \ N}_\beta$,
and do $a/b$ twists/insertions on $n$ of the dimensions you find,
\be Z^{\alpha\ N}_\beta \rightarrow Z^{\alpha\  N-n}_\beta
\mathcal{P}^{\alpha-a\ n}_{\beta+b} \ee

The orbifolded fermion partition function is then
$\sum_{a,b=0}^3 \lambda^a_b$.  Where, \bea\label{hetpart}
(\mbox{HET})\ \ \ \ \ \lambda^a_b &\equiv& \sum_{\alpha,\beta =
0}^{1} \mbox{sgn}_{ab}[\alpha,\beta]^n
Z^{\alpha\ 16-n}_\beta Z^{||\alpha-a|| \ n}_{||\beta + b||} \\
(\mbox{TypeII})\ \ \ \     \lambda^a_b &\equiv& \sum_{\alpha,\beta =
0}^{1} (-)^{\alpha\beta + \alpha +
\beta}(\mbox{sgn}_{ab}[\alpha,\beta])^n Z^{\alpha\ 4-n}_\beta
Z^{||\alpha-a|| \ n}_{||\beta + b||} \ \ . \eea Here $||m|| \equiv
(1-(-)^m)/2$, \ie\ $0$ if $m$ is even and $1$ if $m$ is odd. The
values for $\mbox{sgn}_{ab}[\alpha,\beta]$ can be found using
(\ref{trans2}) and the transformation properties of (\ref{trans1}),
and have been given in Table \ref{sgn}. The factor of
$(-)^{\alpha\beta + \alpha + \beta}$ in the type II case comes from
the relative signs in (\ref{typeiinoorb}).

%
%
%
%

\section{Gradient energy, monodromies and moduli}\label{ap:constant}

Let us now show why the fields $\rho$ and $\tau$ must be fixed under
the action of the monodromy.  It follows from (\ref{sugra}) that if
one allows the moduli of the fiber to vary there is a nontrivial
backreaction on the base.  Such a backreaction is not possible on a
$T^2$.  Let us be more explicit. The metric on the base in complex
coordinates takes the form
\begin{eqnarray}\label{flatmetric}
ds_{base}^2 = \frac{\t \rho_2(z,\zb)}{\t {\tau}_2}|\t \tau d\t
{\theta}_1+ d\t {\theta}_2|^2 \equiv \frac{\t \rho_2(z,\zb)}{\t
\tau_2} |dz|^2 \ ,
\end{eqnarray}
this is just the usual metric written in terms of the moduli $\t
\tau$ and $\t \rho$. It follows that Ricci scalar $R$ is a total
derivative:
\begin{eqnarray}\label{ricci}
R &=& - \nabla^2_{base} \ln \t \rho_2
   = - \frac{\t \tau_2}{\t \rho_2} \d \db \ln \t\rho_2 \\
  &=& - \frac{\t \tau_2}{\t \rho_2} \d \db \ln \rho_2\tau_2
   = - \nabla^2_{base} \ln \rho_2 \tau_2 \ \ .
   \end{eqnarray}
Going from the first line to the second line of (\ref{ricci}) we
have used (\ref{sugra}). The requirement that the base be a
$T^2$ forces the Euler characteristic to vanish;
\begin{eqnarray}\label{chizero}
\chi_{base} = \int_{T^2_{base}} R = 0 \ .
\end{eqnarray}
Since $R$ is a total derivative this is trivially satisfied in
compactifications where $\rho_2$ and $\tau_2$ are single valued.
However, except in the trivial case where $\rho$ and $\tau$ are
constants and take values at fixed points of the monodromy, if
$\tau_2$ or $\rho_2$ undergoes a nontrivial monodromy the surface
term does not vanish. We can now see that in order to satisfy
(\ref{chizero}) and have a nontrivial monodromy, $\rho_2$ and
$\tau_2$ must be fixed under the action of the monodromy.

%
%
%
%

\section{$\zeta$-functions in two sentences}\label{ap:zeta}

The appendices of \cite{adp} review in detail $\vartheta$ and
$\zeta$ functions.  Since the $\zeta$ functions are less familiar
to most readers we define them, and give some of their properties below.
The reader is encouraged to also see the appendix of \cite{adp} which gives
additional properties, as well as specific modular transformations
of the $\zeta$ functions which we use in this paper.

The $\zeta$ functions are a two parameter family of functions
defined by \bea\label{ztadef} \zta \alpha \beta (\tau) \equiv e^{ -
2 \pi i \alpha \beta } {\f {\vt \alpha \beta (0,2 \tau)} {\eta
(\tau)}} =
 {\f 1 {\eta (\tau)}} \sum _{n\in \mathbb{Z}} q^{(n+\alpha)^2}
e^{ 2 \pi i n \beta  } \ . \eea The periodicity properties of
\eqn{ztadef} are \bea \zta {(-\alpha)} {(-\beta)} & = & \zta \alpha
\beta
\nonumber \\
\zta {(\alpha+1)} \beta & = & e^{-2\pi i \beta} \zta \alpha \beta
\nonumber \\
\zta \alpha {(\beta+1)} & = &
 \zta \alpha \beta ,
\eea and under modular transformations \bea \zta \alpha  \beta
(\tau+1) & = & e^{-i\pi/12} e^{ 2 \pi i \alpha ^2} \zta \alpha
{(\beta +2 \alpha)} (\tau)
\nonumber \\
\zta \alpha  \beta  (-{\f 1 \tau}) & = & {\f 1 {\sqrt{2}}}
e^{-2\pi i \alpha \beta} \biggl( \zta {(-\f {\beta} 2)} {2 \alpha}
(\tau) + e^{ 2 \pi i \alpha} \zta {(\f 1 2 -{\f \beta 2})} {2
\alpha}(\tau) \biggr)\ . \eea

\acknowledgments
We would like to thank Peter Bouwknegt, Eric Verlinde, Nick Jones and Brian Wecht for 
useful discussions.  BW is supported by NSF grant OISE-0402111. 
AF has been supported through a grant from the Australian Research 
Council.
\bibliography{fw}

\end{document}